%% file: digital_twin_csi_AA2.tex
\documentclass[conference]{IEEEtran}
\usepackage[top=0.7in, bottom=0.7in, left=0.62in, right=0.62in]{geometry}

\input{input.tex}
\usepackage{graphicx}
\usepackage{epstopdf}
\usepackage{enumerate}
\usepackage{algorithm}
\usepackage{amsmath}
\usepackage{mathrsfs}
\usepackage{float}
\usepackage{color}
\usepackage{makeidx}
\usepackage{bm}
\usepackage{cleveref}
\usepackage{url}
\usepackage{steinmetz}
\usepackage{varwidth}
\usepackage{soul}
\usepackage{bbm}
\usepackage{empheq}
\usepackage{mathtools}
\usepackage{cite}
\usepackage{balance}

\newcommand{\sref}[1]{{Section}~\ref{#1}}

\def \rm {\mathrm}

\usepackage{amsfonts}
\usepackage{booktabs}
\usepackage{siunitx}

\begin{document}
	\title{Digital Twin Aided Massive MIMO:\\CSI Compression and Feedback}
	\author{Shuaifeng Jiang and Ahmed Alkhateeb \\Arizona State University - Emails: \{s.jiang, alkhateeb\}@asu.edu}

\maketitle
\begin{abstract}
	Deep learning (DL) approaches have demonstrated high performance in compressing and reconstructing the channel state information (CSI) and reducing the CSI feedback overhead in massive MIMO systems. One key challenge, however, with the DL approaches is the demand for extensive training data. Collecting this real-world CSI data incurs significant overhead that hinders the DL approaches from scaling to a large number of communication sites. To address this challenge, we propose a novel direction that utilizes site-specific \textit{digital twins} to aid the training of DL models. The proposed digital twin approach generates  site-specific synthetic CSI data from the EM 3D model and ray tracing, which can then be used to train the DL model without real-world data collection. To further improve the performance, we adopt online data selection to refine the DL model training with a small real-world CSI dataset. Results show that a DL model trained solely on the digital twin data can achieve high performance when tested in a real-world deployment. Further, leveraging domain adaptation techniques, the proposed approach requires orders of magnitude less real-world data to approach the same performance of the model trained completely on a real-world CSI dataset. 
\end{abstract}

\begin{IEEEkeywords}
	CSI feedback, digital twin, deep learning, massive MIMO.
\end{IEEEkeywords}

\section{Introduction}\label{Introduction}
Multiple-input and Multiple-Output (MIMO) communication	gains have been widely recognized as a critical feature of advanced wireless communication systems. By adopting a large number of antennas, massive MIMO can provide promising gains in spectral, spatial, and energy efficiency. Fully harvesting these gains, however, requires accurate information about the communication channels at the base station (BS). The downlink channel acquisition process typically consists of three steps. (i) First, the BS sends downlink pilot signals to the user equipment (UE). (ii) After that, the UE estimates the downlink channel state information (CSI) from the received pilot signals. (iii) The UE feeds the estimated downlink CSI back to the BS through the uplink control channel.
However, with the classical CSI quantization  approaches, the CSI feedback grows quickly with the number of antennas, making it difficult for MIMO system to continue scaling their gains with more antennas.

Recently, deep learning (DL) approaches have been widely investigated to reduce the CSI feedback overhead \cite{wen2018deep,wang2018deep,guo2020convolutional,ji2021clnet}. The DL CSI feedback approaches typically employ an autoencoder neural network (NN) consisting of an encoder and a decoder. In particular, the UE first uses the encoder to compress the full CSI to a lower-dimension representation, and then feeds back the compressed CSI. The BS applies the decoder to reconstruct the full CSI from the compressed one. In \cite{wen2018deep}, the author proposed to employ DL to solve the CSI feedback problem for the first time. \cite{wang2018deep} leveraged the temporal structure of wireless channels and developed DL CSI feedback solutions for time-varying channels. \cite{guo2020convolutional} proposed a multiple-rate NN to compress and quantize the CSI. In \cite{ji2021clnet}, the author proposed a novel NN architecture for the CSI feedback task considering the complex-value CSI. In \cite{cui2022unsupervised}, the author presented an unsupervised online learning approach to deal with channel statistics changes. However, one key challenge with the prior work is the need for extensive real-world training CSI data. From a machine learning perspective, the training data distribution needs to align well with the test data distribution for the DL model to achieve high performance. Since the CSI distribution is closely related to the communication environment, scaling the DL approaches to a large number of communication sites requires a large real-world CSI data collection overhead.

In this paper, we propose to leverage  site-specific digital twins \cite{alkhateeb2023real} to reduce the real-world CSI data collection overhead for the DL CSI feedback task. The contribution of this paper can be summarized as follows. (i) We propose to employ a site-specific digital twin to aid the training of the DL model and reduce the real-world CSI collection overhead. (ii) We propose a data selection and model refinement approach to refine the DL model. The proposed approach can compensate for the mismatch between the real-world and digital twin CSI distributions and improve the DL performance with a small number of real-world data.

\section{System and Channel Models}\label{System and Problem Formulation}
We consider a frequency division duplex (FDD) single-cell system where a base station (BS) with $N_t$ antennas is communicating with a single user equipment (UE) with one antenna. The system adopts the OFDM modulation with $K$ subcarriers. Let $x_k \in \bbC$ denote the downlink complex symbol transmitted on the $k$-th $(k\in\{1,\hdots,K\})$ subcarrier with unit average power $\bbE[x_k^\textrm{H}x_k]=1$. The received signal at the $k$-th subcarrier can then be written as
\begin{align}
	y_k = \bh_k^\textrm{H}\bff_k x_k + n_k,
\end{align}
where $\bh_k \in \bbC^{N_t\times 1}$ and $\bff_k \in \bbC^{N_t\times 1}$ denote the downlink channel vector and precoding vector of the $k$-th subcarrier, respectively. $n_k \in \bbC$ is the additive Gaussian noise with zero mean and variance $\sigma^2_n$. For the channel $\bh$, we adopt a block-fading wideband geometric channel model. The delay-d channel vector $\bh_d$ can be written as
\begin{align}\label{eq:delayd_channel}
	\bh_d = \sum_{l=1}^{L}\alpha_lp(dT_S-\tau_l)\ba(\phi_l, \theta_l),
\end{align}
where $p(\tau)$ denotes the pulse shaping function that represents a $T_S$-spaced signaling evaluated at $\tau$ seconds. The channel consists of $L$ paths, and $\alpha_l$, $\tau_l$, $\phi_l$, and $\theta_l$ denotes the complex gain, propagation delay, azimuth and elevation angles of departure (AoD) of the $l$-th path, respectively. $\ba(\phi_l, \theta_l)$ is the BS array response, which depends on the array geometry. For simplicity, we consider a uniform linear array (ULA) with half-wavelength antenna spacing. The frequency domain channel vector on the $k$-th subcarrier can then be written as
\begin{align}\label{eq:frequency_channel}
	\bh_k = \sum_{d=0}^{D-1}\bh_d \exp\left(-j\frac{2\pi k}{K}d\right),
\end{align}
where $D$ denotes the maximum delay of the channel. The sum rate of the downlink data transmission can then be written as
\begin{equation}
	R = \sum_{k=1}^K\log_2\left(1+\frac{P_t}{KN_t\sigma_n^2}  \left|\bh_k^\textrm{H}\,\bff_k\right|^2\right),
\end{equation}
where $P_t$ is the total transmit power of the BS.

Let $\bH = [\bh_1, \hdots, \bh_K]^\textrm{H} \in \bbC^{K \times N_t}$ denote the downlink CSI matrix obtained by stacking the $K$ channel vectors. To achieve high sum rates $R$, the BS requires accurate downlink CSI $\bH$ to optimize the precoding vectors $\bff_k$. In FDD systems, the BS relies on the downlink training and uplink feedback to obtain the downlink CSI. In particular, the BS first sends pre-defined downlink pilots to the UE. The UE then estimates the CSI based on the received pilots, and returns the estimated CSI via the uplink feedback channel. Let $\widehat{\bH}=[\widehat{\bh}_1, \hdots, \widehat{\bh}_K]^\textrm{H} \in \bbC^{K \times N_t}$ denote the CSI feedback received at the BS. Since we primarily focus on CSI compression and recovery, we assume that the UE estimates the CSI perfectly and the uplink feedback channel is lossless. Since we adopt a block-fading channel model, the CSI remains constant within the downlink pilot training, uplink feedback, and downlink data transmission periods.
\section{Deep Learning CSI Feedback}
The UE needs to feed back the estimated CSI matrix to the BS. However, since the dimensions of the CSI matrix grow proportionally with the number of antennas and subcarriers, directly sending the CSI matrix leads to large wireless resource overhead. Given the limited channel coherence time, this CSI feedback overhead can significantly degrade the system performance. To reduce the feedback overhead, the CSI is typically first compressed at the UE. Let $f_\textrm{enc}(\cdot)$ denote the CSI compression function (encoder). The compressed CSI $\bg \in \bbC^{M \times 1}$ can be written as $\bg = f_\textrm{enc}(\bH)$,
where the dimension of the compressed CSI is smaller than the CSI matrix, \textit{i.e.}, $M< N_tK$. Then, the compressed CSI $\bg$ is sent to the BS with reduced feedback overhead. The BS then recovers the CSI matrix $\widehat{\bH}$ from the compressed CSI $\bg$, which can be written as $\widehat{\bH} = f_\textrm{dec}(\bg) = f_\textrm{dec}\big(f_\textrm{enc}(\bH)\big)$,
where $f_\textrm{dec}(\cdot)$ denotes the CSI recovery function (decoder).

This paper aims to design CSI compression and recovery functions using DL models so that the CSI recovered at the BS closely matches the estimated CSI at the UE for a given channel distribution \mbox{$\bH \sim \cH$}. To evaluate the error between the estimated CSI and the recovered CSI, we employ the normalized mean square error (NMSE) given by
\begin{equation}\label{eq:nmse}
	\mathsf{NMSE}(\bH, \widehat{\bH}) = \frac{\|\bH - \widehat{\bH}\|_F^2}{\|\bH\|_F^2},
\end{equation}
where $\|\cdot\|_F$ denotes the Frobenius norm.
Let $f_\textrm{enc}(;\Theta_\textrm{enc})$ and $f_\textrm{enc}(;\Theta_\textrm{dec})$ denote the DL models for the CSI compression and recovery functions, with $\Theta_\textrm{enc}$ and $\Theta_\textrm{dec}$ denoting the trainable parameters. Our objective is to design DL models that minimize the channel recovery NMSE. The optimal DL models can be written as
\begin{align}\label{eq:optimal_ml_model}
	& f^\star_\textrm{enc}(;\Theta_\textrm{enc}^\star), \, f^\star_\textrm{dec}(;\Theta_\textrm{dec}^\star) \nn\\
	=\,& \underset{\substack{f_\textrm{enc}(;\Theta_\textrm{enc})\\ f_\textrm{dec}(;\Theta_\textrm{dec})}}{\arg\min} \ \bbE_{\bH \sim \cH}\bigg\{  \mathsf{NMSE}\Big[\bH, f_\textrm{dec}\big(f_\textrm{enc}(\bH)\big)\Big]\bigg\}.
\end{align}

In practice, the exact CSI distribution is approximated by a dataset of the CSI matrices. Let \mbox{$\cD = \{\bH_1, \hdots, \bH_{D}\}$} denote the CSI dataset consisting of $D$ data points drawn from $\cH$, the DL models can then be trained in an end-to-end manner by minimizing the following loss function:
\begin{align}\label{eq:ml_loss}
	L(\Theta_\textrm{enc}, \Theta_\textrm{dec}, \cD) = \frac{1}{D}\sum_{d=1}^{D} \mathsf{NMSE}\Big[\bH_d, f_\textrm{dec}\big(f_\textrm{enc}(\bH_d)\big)\Big].
\end{align}

\begin{figure*}[t]
	\centering
	\includegraphics[width=0.9\linewidth]{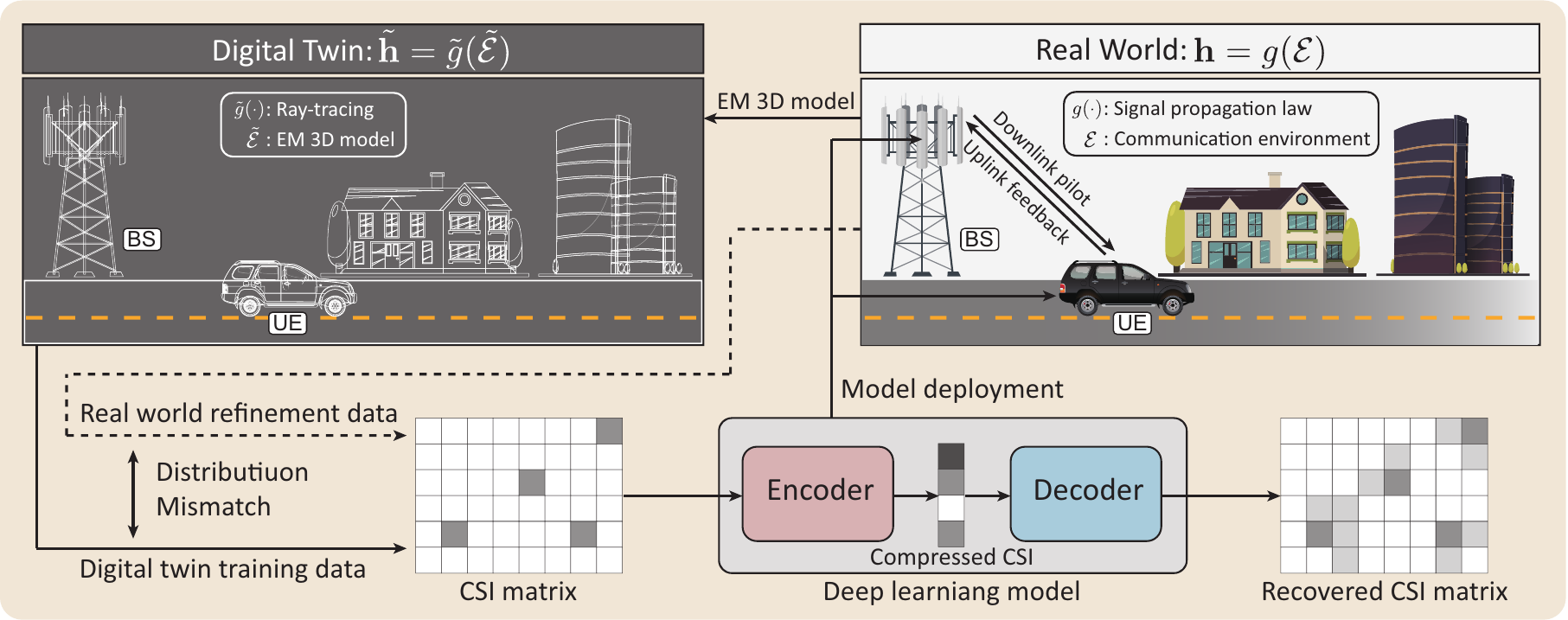}
	\caption{This figure shows the key idea of utilizing a digital twin to train the DL CSI model. A small amount of real-world data can then be used to compensate for the mismatch between the digital twin and real-world CSI data, and refine the DL model to achieve higher performance.}
	\label{fig:key_idea}
\end{figure*}
\section{Digital Twin Aided Deep Learning CSI Feedback}
One challenge with the DL approach is the demand on real-world training data. The DL model typically requires a dataset that captures the full diversity of the CSI distribution to ensure robust performance. Note that the CSI distribution is closely related to the communication environment and varies for different sites. The consequent extensive real-world CSI data collection overhead is a significant barrier to scaling the DL CSI feedback across a large number of communication sites. Next, we introduce the key idea of leveraging digital twin to reduce or even eliminate this real-world data collection overhead.

\subsection{Key Idea}
The real-world communication channel is largely determined by: (i) the positions, orientations, dynamics, shapes, and materials of the BS, the UE, and other objects (reflectors/scatterers) in the communication environment denoted by $\cE$, and (ii) the wireless signal propagation law denoted by $g(\cdot)$. Given the communication environment and wireless signal propagation law, the communication channel can be obtained by
\begin{align}
	\bH = g(\cE).
\end{align}
However, the precise communication environment $\cE$ is difficult to obtain, and the exact expression of the signal propagation law $g(\cdot)$ is difficult to characterize in complex environments.

To that end, we propose to construct a site-specific digital twin that approximates the communication environment $\cE$ and signal propagation law $g(\cdot)$ \cite{alkhateeb2023real}. In particular, the digital twin approximates the communication environment $\cE$ with EM 3D models and the signal propagation law $g(\cdot)$ with ray tracing.

\textbf{EM 3D model.} The EM 3D model  $\widetilde{\bE}$ contains (estimated) information about the positions, orientations, dynamics, shapes, and materials of the BS, the UE, and other surrounding objects (reflectors/scatterers). The 3D model information can be obtained from architectural design drawings, remote sensing, and field measurements.

\textbf{Ray tracing.} The ray tracing $\widetilde{g}(\cdot)$ models the propagation paths between each transmit-receive antenna pair based on the geometry and material data from 3D models considering multiple propagation effects. For each path, ray tracing generates path parameters, including path gain, propagation delay, and propagation angles, which can synthesize channels in a way spatially consistent with the geometry. 

The site-specific digital twin can generate synthetic CSI $\widetilde{\bH} = \widetilde{g}(\widetilde{\bE})$ that approximates the real-world CSI. If the synthetic data closely resemble the real-world data, we can leverage it in the training of the CSI feedback model, and reduce/eliminate the real-world CSI data collection.

\subsection{Problem Formulation}
Let $\cH$ and $\widetilde{\cH}$ denote the CSI distributions in the real world and the digital twin, respectively. We aim to leverage the digital twin CSI distribution to obtain a DL model that can achieve similar performance as the DL model trained on the real-world CSI distribution. Let $\tilde{f}_\textrm{enc}(;\widetilde{\Theta}_\textrm{enc})$ and $\tilde{f}_\textrm{dec}(;\Theta_\textrm{dec})$ denote the DL models obtained using the digital twin CSI distribution $\widetilde{H}$, the objective can be mathematically formulated by
\begin{align}\label{eq:objective}
	\underset{\substack{\tilde{f}_\textrm{enc}(;\widetilde{\Theta}_\textrm{enc})\\ \tilde{f}_\textrm{dec}(;\widetilde{\Theta}_\textrm{dec})}}{\min} \left| L(\widetilde{\Theta}_\textrm{enc}, \widetilde{\Theta}_\textrm{dec}, \cH)-L(\Theta_\textrm{enc}^\star, \Theta_\textrm{dec}^\star, \cH)\right|.
\end{align}

\section{Proposed Solutions}
\subsection{Direct Generalization} \label{sec:Direct Generalization}
Training a DL model on the digital twin synthetic data to achieve high performance on the real-world data is a domain adaptation problem from an ML perspective \cite{zhao2020review}. In this case, the digital twin synthetic data distribution is the source domain, and the real-world data distribution is the target domain. The domain adaptation bounds have been investigated in \cite{mansour2009domain}. In \cite{mansour2009domain}, the authors showed that, for any mapping function, \textit{i.e.} the CSI compression and recovery models $\widetilde{f}_\textrm{enc}(;\widetilde{\Theta}_\textrm{enc})$ and $\widetilde{f}_\textrm{enc}(;\widetilde{\Theta}_\textrm{dec})$ in our context, the following bound hold:
\begin{align}\label{eq:bound}
	&\left| L(\widetilde{\Theta}_\textrm{enc}, \widetilde{\Theta}_\textrm{dec}, \cH)-L(\Theta_\textrm{enc}^\star, \Theta_\textrm{dec}^\star, \cH)\right| \nonumber\\
	\leq \ &L(\widetilde{\Theta}_\textrm{enc}, \widetilde{\Theta}_\textrm{dec}, \widetilde{\cH}) + \mathrm{disc}(\cH, \widetilde{\cH}) + \epsilon,
\end{align}
where $\mathrm{disc}(\cH, \widetilde{\cH}) > 0 $ denotes the discrepancy distance between the two distributions $\cH$ and $\widetilde{\cH}$. $\epsilon > 0$ is a constant given $\cH$, $\widetilde{\cH}$, and the function class of $f_\textrm{enc}$ and $f_\textrm{dec}$.
Since $\mathrm{disc}(\cH, \widetilde{\cH})$ and $\epsilon$ are both constants given $\cH$ and $\widetilde{\cH}$, the upper bound of \eqref{eq:objective} can be minimized by minimizing $L(\widetilde{\Theta}_\textrm{enc}, \widetilde{\Theta}_\textrm{dec}, \widetilde{\cH})$, which is the DL model's loss on the digital twin synthetic data. To that end, the DL model can be obtained by minimizing the following loss function:
\begin{align}\label{eq:ml_loss2}
	L(\widetilde{\Theta}_\textrm{enc}, \widetilde{\Theta}_\textrm{dec}, \widetilde{\cD}) = \frac{1}{\widetilde{D}}\sum_{d=1}^{\widetilde{D}} NMSE\Big[\widetilde{\bH}_d, \widetilde{f}_\textrm{dec}\big(\widetilde{f}_\textrm{enc}(\widetilde{\bH}_d)\big)\Big],
\end{align}
where $\widetilde{\cD}$ is a dataset consisting of $\widetilde{D}$ CSI matrices sampled from the digital twin $\widetilde{\cH}$. Note that \eqref{eq:bound} and \eqref{eq:ml_loss2} align with the intuition: When the digital twin data closely resembles the real-world data, the DL model directly trained on the digital twin data can achieve high performance on the real-world data.

\subsection{Model Refinement}\label{sec:Model Refinement}
In real-world deployments, it is difficult to construct a digital twin that perfectly matches the real-world scenario. Due to the impairments in the EM 3D model and ray tracing, the digital twin synthetic data distribution can differ from the real-world data distribution. That is, the impairments in the digital twin leads to a larger discrepancy distance $\mathrm{disc}(\cH, \widetilde{\cH})$. To alleviate the effect of the impairments in the digital twin, a small amount of real-world data can be utilized to refine the DL model trained on the digital twin data. Let $\cD_r$ denote a small real-world CSI dataset for refining the DL model that is previously trained on $\widetilde{\cD}$. In this work, we investigate the following two model refinement approaches.

\textbf{Naive Fine-Tuning.} A straightforward way to refine the DL model on the real-world dataset $\cD_r$ is the naive fine-tuning. After the DL model is trained on the digital twin dataset $\widetilde{\cD}$, this approach refines the DL model by minimizing the loss function in \eqref{eq:ml_loss} on the dataset $\widetilde{\cD}$.

\textbf{Rehearsal.} The naive fine-tuning may suffer from overfitting when the refining dataset is small. Moreover, while the DL model fits to the refining dataset, it may lose the generalizable patterns learned from the digital twin synthetic dataset, which leads to the catastrophic forgetting effect \cite{robins1995catastrophic}. To compact the overfitting and catastrophic forgetting, we adopt the rehearsal approach in the model refinement process \cite{robins1995catastrophic}. Rehearsal is the relearning of the previously learned data samples while the new training data samples are introduced. By also training on the learned data samples in the refinement process, rehearsal strengthens or preserves the previously learned information from the disruption by new data samples. Moreover, adding the learned data samples in the refining process also increases the size of the refining dataset and mitigates overfitting. The rehearsal refines the DL model by minimizing the loss function in \eqref{eq:ml_loss} on the combination of datasets $\widetilde{\cD}\cup \cD_r$.

\textbf{System Operations.} To conduct the model refinement, we assume that the system operates with the following three phases: (i) offline pre-training, (ii) online refinement data collection, and (iii) model refinement and update.
\begin{itemize}
	\item \textbf{Offline Pre-Training.} First, the digital twin is constructed at the infrastructure side. The digital twin is used to generate training data to train the DL CSI feedback model as detailed in \sref{sec:Direct Generalization}. Then, the trained DL models (encoder and decoder) are sent to the UEs.
	
	\item \textbf{Online Refinement Data Collection.} After the DL model is pre-trained, the BS and the UE start to use the DL model for the CSI compression and recovery tasks. In addition, the UE can choose to feedback a small number of full CSI matrices to the BS for the model refinement.
	
	\item \textbf{Model refinement and update.} The BS collects the full CSI matrices fed back from the UEs. Once a sufficient number of full CSI matrices are received, the BS can construct the refinement dataset $\cD_r$, and refine the DL model. After that, the updated model is sent to the UEs and used for the CSI compression and recovery tasks. Other than compensating for the impairment in the digital twin data, the system can also leverage this refinement data collection and model update operations to maintain the DL model when, \textit{e.g.}, the environment changes.
\end{itemize}

\subsection{Refinement Data Selection} \label{sec:Refinement Data Selection}
In the online refinement data collection stage, the UE has the opportunity to select which full CSI matrices to feedback to the BS, which can be optimized to improve the model refinement performance. As a baseline data selection method, the UE can randomly select some CSI matrices to feed back to the BS. However, this random data selection can be inefficient. When the digital twin data approximates the real-world data relatively well, the randomly selected CSI matrices are likely to have a high correlation with the digital twin data. In this case, these randomly selected CSI barely provide new information about the real-world channel distribution and, therefore, is less effective for the model refinement. To that end, we propose a heuristic approach for the UE to select the feedback CSI matrices more efficiently. In particular, after the UE estimates the CSI matrix $\bH$, it first uses the encoder and decoder DL model to compress and recover the CSI matrix. After that, the UE calculates the reconstruction NMSE between the original CSI matrix $\bH$ and the recovered CSI matrix $f_\mathrm{dec}(f_\mathrm{enc}(\bH))$ using \eqref{eq:nmse}. The UE then feeds back the CSI matrices whose NMSE is larger than a threshold $\eta$.

\subsection{Deep Learning Model Design}\label{sec:Deep Learning Model Design}
\textbf{Data Pre-Processing.} We follow the common practice of transforming the frequency-antenna domain channel $\bH$ to the delay-angular domain to obtain a sparse representation \cite{guo2020convolutional}. The delay-angular domain channel matrix $\bG$ can be obtained by
\begin{equation}
	\bG = \bF_\textrm{d} \bH \bF^\textrm{H}_\textrm{a},
\end{equation}
where $\bF_\textrm{d}$ and $\bF_\textrm{a}$ are $K$-dimensional and $N_t$-dimensional DFT matrices, respectively. Further, in the delay domain, only the first several rows of $\bG$ have non-zero elements due to the limited delay spread. Therefore, the $\bG$ is truncated to $\bG_\textrm{trunc}$ to keep the first 32 rows and remove the others. Lastly, we normalize $\bG_\textrm{trunc}$ to $\bG_\textrm{trunc}$ with unit Frobenius norm:
\begin{equation}\label{eq:norm}
	\bG_\textrm{norm} = \frac{\bG_\textrm{trunc}}{\|\bG_\textrm{trunc}\|_F}.
\end{equation}

The input data and the ground-truth output data of the CSI feedback DL model both take the format of the normalized delay-angular CSI matrix $\bG_\textrm{norm}$. Note that this normalization does not affect the downlink precoding design at the BS.

\textbf{NN Architecture.} We adopt a NN architecture similar to the CSINet+ \cite{guo2020convolutional} with the following modifications. (i) We replace the sigmoid activation function before the RefineNet Blocks with the tanh activation function. The output activation function is designed to work as an initial estimate of the $\bG_\textrm{norm}$. With the normalization in \eqref{eq:norm}, the elements of the normalized CSI matrix $\bG_\textrm{norm}$ lies in $[-1,1]$. Therefore, we employ the tanh activation function to regularize the elements in the initial estimate into the same range. (ii) After the final layer in the decoder in CSINet+ \cite{guo2020convolutional}, we added a normalization layer similar to \eqref{eq:norm} that normalizes the Frobenius norm of the output recovered angular-delay domain CSI matrix. We found the input and output normalization layers stabilize and accelerate the training process of the DL model. We adopt a CSI compression rate of 64, and the dimension of the compressed CSI is $M=32$.

\section{Simulation Setup}
This section explains the simulation scenario and dataset.

\subsection{Scenario Setup}
\begin{figure}[t]
\centering
\includegraphics[width=0.8\linewidth]{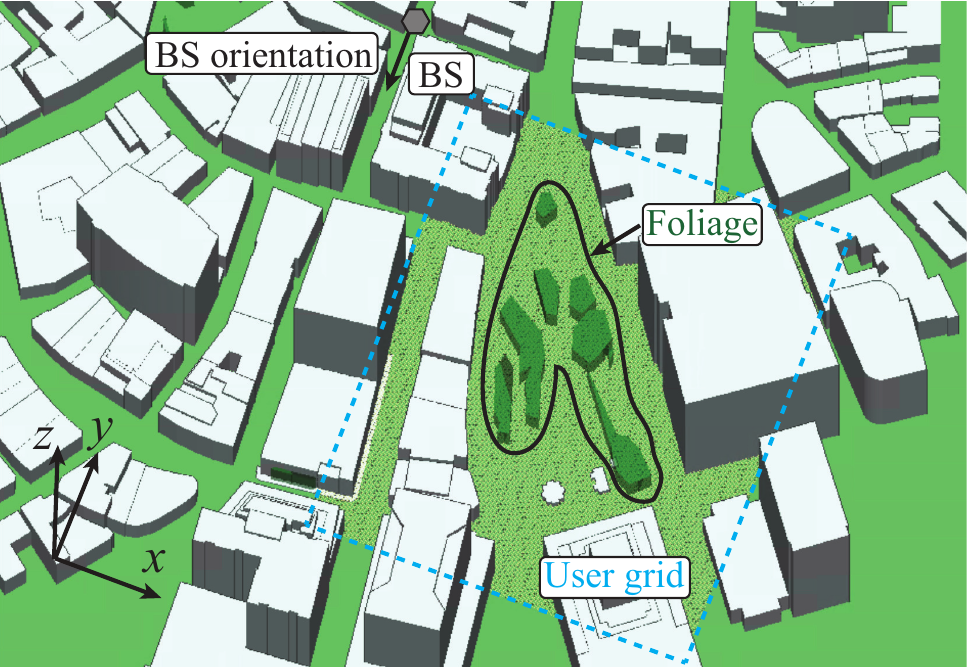}
\caption{This figure shows the geometry layout of the target scenario, which reflects a real-world section in Boston City. The BS is located in a vertical street, and the service area is annotated in blue.} 
\label{fig:scenario_boston}
\end{figure}

\begin{figure}[t]
\centering
\includegraphics[width=0.8\linewidth]{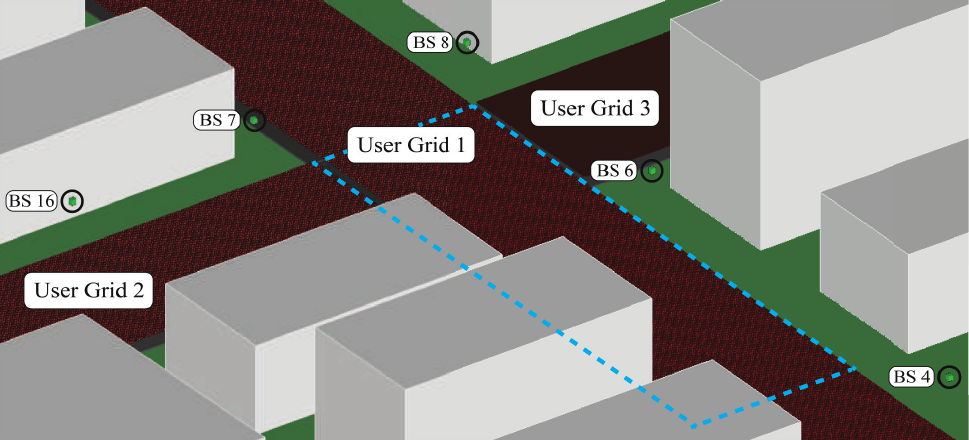}
\caption{This figure shows the geometry layout of the baseline scenario, which adopts the O1 scenario from the DeepMIMO dataset \cite{deepmimo}. We set ``BS 6" as the BS, and activate rows 1150 to 1650 as the service area (annotated in blue).}
\label{fig:scenario_o1}
\end{figure}

We employ three ray tracing scenarios in our evaluation: (i) the target scenario, (ii) the digital twin scenario, and (iii) the baseline scenario. 

\textbf{Target Scenario.} The target scenario is considered the ``real-world" scenario in our evaluation. \figref{fig:scenario_boston} shows the bird's view of the target scenario, which reflects a real-world urban section in Boston city. The target scenario incorporates a BS, a service area, and multiple building and foliage objects. The BS utilizes a 32-element ULA antenna, with a height of 15 meters, oriented towards the negative y-axis. The service area (annotated in blue) covers where the UE can appear. This area is 200 meters by 230 meters, and the height of the UE is  2 meters. The service area is discretized into a user grid of multiple UE positions with a spacing of 0.37 meters. 

\textbf{Digital Twin Scenario.} In real-world implementation, the digital twin may not be able to model all objects in the environment accurately. For instance, the digital twin may more accurately model buildings. Nonetheless, while they significantly influence the wireless channel, other elements, such as the foliage, become more challenging to model due to their seasonal variations. For the digital twin scenario, we adopt the same geometry layout as the target scenario with the foliage objects neglected. This way, we capture the similarity between the target and digital twin scenarios, and emulate the modeling impairment in the digital twin. 

\textbf{Baseline Scenario.} The proposed digital twin-aided CSI feedback approach utilizes the data generated by the site-specific digital twin. To evaluate the efficacy of the site-specific digital twin CSI data, we are interested in a comparative study with the CSI data generated from an entirely distinct geometry layout. For that, we employ the O1 scenario from the DeepMIMO dataset \cite{deepmimo}. As shown in \figref{fig:scenario_o1}, we select the ``BS 6" as the BS and choose rows 1150 to 1650 of ``User Grid 1" as the discretized service area (annotated in blue).

\subsection{Dataset Generation} 
We first run accurate 3D ray tracing between the BSs and all UE positions in the target, digital twin, and baseline scenarios. The ray tracing tracks the propagation paths between the BS and each UE position, respecting the geometry layout. For each propagation path, it produces the path parameters: complex gain $\alpha_l$, the propagation delay $\tau_l$, azimuth AoD $\phi_l$, and elevation AoD $\theta_l$. We assume that the BS serves the UE at the 3.5 GHz band with 256 subcarriers. From the path parameters, we then generate the CSI matrices $\bH$ between the BS and the UE positions using \eqref{eq:delayd_channel} and \eqref{eq:frequency_channel} with the DeepMIMO channel generator \cite{deepmimo}. After that, we apply the pre-processing in \sref{sec:Deep Learning Model Design} to convert the $\bH$ to the delay-angular domain representation $\bG_\textrm{norm}$. For each scenario, we split the CSI matrices into $80\%$ and $20\%$ to obtain the training and test datasets. 

\section{Evaluation Results}\label{sec:Evaluation Results}
In this section, we evaluate the performance of the proposed digital twin-aided CSI feedback approaches. 
\subsection{Direct Generalization}
We first investigate the performance of the direct generalization approach. \figref{fig:data_size} shows the CSI reconstruction NMSE obtained by training the DL model on different numbers of digital twin data points, and testing the DL model on the target data. For the benchmarks, we train the DL model on the same numbers of data points from the baseline or the target scenario. All models are tested on the unseen \textit{target} scenario data. As can be seen from \figref{fig:data_size}, among the three direct generalization approaches, training on the target scenario data leads to the best NMSE performance because the training data and the testing data are from the same distribution.
The DL model trained on baseline scenario data, however, cannot generalize to the target scenario data. Moreover, increasing the number of training data points cannot improve the NMSE performance on the target data. The baseline scenario has a very different geometry layout than the target scenario. The consequent much different CSI distribution limits the generalization performance. In contrast, \textbf{the digital twin captures the \textit{site-specific} geometry layout in the EM 3D model, which can generate CSI data that approximate the target (``real-world") data.} This enables the DL model to generalize well on the target scenario data. For instance, after trained on 5120 digital twin data points, the DL model achieves $-17$ dB NMSE on the target data. \textbf{Note that this approach does not utilize any target data in the training process and, therefore, does not require any real-world CSI data collection.} The performance gap between training on the target and digital twin data becomes larger as the training dataset size increases. When the training dataset is small, the performance is mainly limited by the quantity of the available training data. As the training dataset size increases, the performance becomes more bottlenecked by the mismatch between the digital twin and the target (``real-world") scenario.

\subsection{Model Refinement}
\begin{figure}[t]
\centering
\includegraphics[width=0.85\linewidth]{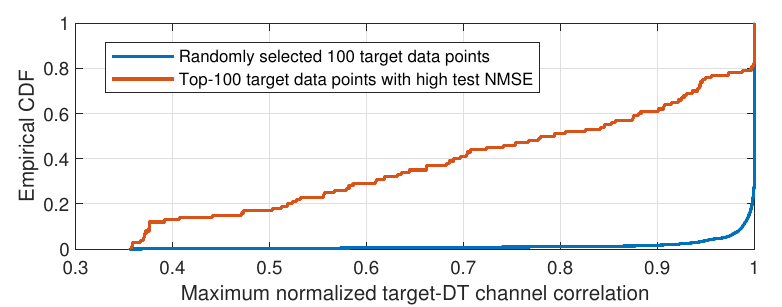}
\caption{This figure shows the empirical CDF of the normalized correlation between the target and digital CSI matrices.}
\label{fig:correlation}
\end{figure}

Here, we investigate the performance of the refinement data selection and model refinement approaches. First, in the offline pre-training phase, we train a DL model on 5120 data points from the digital twin scenario. This pre-trained model is used to select the target CSI data for model refinement. In particular, this model is applied to compress and recover all the CSI matrices in the training dataset from the target scenario and obtain the reconstruction NMSE for each of these CSI matrices.
To evaluate the proposed data selection approach, we calculate the maximum normalized correlation between each target CSI matrix and all the digital twin CSI matrices. Let $\widetilde{\cD} = \{\widetilde{\bH}_1, \hdots, \widetilde{\bH}_{\widetilde{D}}\}$ denote the dataset containing the $\widetilde{D}$ digital twin CSI matrices. The maximum correlation calculated from the target CSI matrix $\widetilde{\bH}$ can be written as
\begin{equation}
\max_{i=1,\hdots,\widetilde{D}} \frac{|u(\bH)^\textrm{H}u(\widetilde{\bH}_i)|}{\|u(\bH)\|_{2}\|u(\widetilde{\bH}_i)\|_{2}},
\end{equation}
where $u(\bH)$ flattens the CSI matrix $\bH$ to one-dimensional, and $\|\cdot\|_{2}$ denote the $L_2$ vector norm. When the maximum correlation is high (close to one), there is a CSI matrix in the digital twin dataset similar to the target CSI matrix. Then, this target CSI matrix does not provide much new information given the digital twin data and is less effective for model refinement. In \figref{fig:correlation}, we show the empirical cumulative distribution function (CDF) for the normalized correlation calculated from the target CSI matrices with the top-100 highest reconstruction NMSE. As a benchmark, we present the same plot with 100 randomly select CSI matrices. It shows that the target CSI matrices with high reconstruction NMSE lead to a significantly smaller normalized correlation. Therefore, these target CSI matrices better represent the mismatch between the target and digital twin data. This demonstrates the efficacy of the proposed refinement data selection approach.

\begin{figure}[t]
\centering
\includegraphics[width=0.85\linewidth]{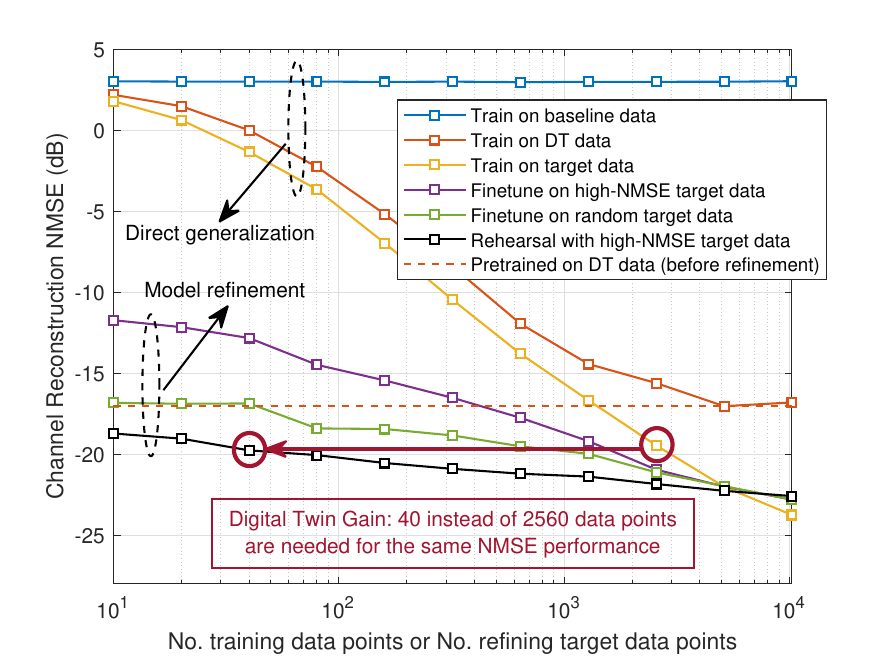}
\caption{This figure shows the NMSE performance of the direct generalization and three model refinement approaches. All NMSE performance is evaluated on the target data unseen in the training and refining.}
\label{fig:data_size}
\end{figure}

\figref{fig:data_size} presents the NMSE performance of refining the pre-trained model using three approaches. For comparison, we also show the performance of the pre-trained model without refinement. When the model is fine-tuned on a small number of randomly selected target data, the NMSE performance does not improve since these target data are unlikely to capture the mismatch between the target and digital twin CSI distribution. When the model is fine-tuned on a small number of high-NMSE target data, the performance becomes even worse than the pre-trained model. While the high-NMSE target data can capture the mismatch between the target and digital twin, they differ significantly from the digital twin data used in the pre-training. Therefore, directly fine-tuning the pre-trained model on this data can disrupt the previously learned patterns of the model. In contrast, refining the model using rehearsal can efficiently calibrate the mismatch between the digital twin and the target scenarios while preserving the previously learned patterns. \textbf{This enables further improving the pre-trained model with a small number of target data.} With only 40 target data points, the NMSE performance can be further improved to $-20$ dB.


\section{Conclusion}\label{Conclusion}
This paper investigates the DL CSI feedback for massive MIMO systems. To reduce the real-world CSI data collection overhead for the DL training process, we propose a novel direction that utilizes a site-specific digital twin to aid the training of the DL model. In particular, the communication infrastructure can leverage the EM 3D model to approximate the real-world communication environment. By applying ray tracing on the EM 3D model, a digital twin of the real-world communication system can be constructed.
The digital twin can be used to generate synthetic CSI data that closely resemble the real-world CSI distribution and train DL models.
Moreover, to compensate for the impairments in the digital twin and improve the DL performance, we  adopt online data selection  and leverage domain adaptation techniques to refine the DL model training with a small real-world CSI dataset. 
Results show that training the model solely on the digital twin data can achieve high CSI reconstruction performance on the target real-world scenario. Moreover, the proposed domain-adaptation-based refinement solution can approach the upper bound with orders of magnitude less real-world CSI data points.


\end{document}

%% file: input.tex
\usepackage{amsfonts}
\usepackage{times}
\usepackage{graphicx}
\usepackage{epstopdf}
\usepackage{latexsym}
\usepackage{dsfont}
\usepackage{amssymb}
\usepackage{amsmath}
\usepackage{cite}
\usepackage{verbatim}

\newcommand{\figref}[1]{{Fig.}~\ref{#1}}

\def\bb0{{\mathbb{0}}}

\def\ba{{\mathbf{a}}}
\def\bb{{\mathbf{b}}}

\def\bff{{\mathbf{f}}}
\def\bg{{\mathbf{g}}}
\def\bh{{\mathbf{h}}}

\def\b0{{\mathbf{0}}}

\def\bE{{\mathbf{E}}}
\def\bF{{\mathbf{F}}}
\def\bG{{\mathbf{G}}}
\def\bH{{\mathbf{H}}}

\def\bbC{{\mathbb{C}}}

\def\bbE{{\mathbb{E}}}

\def\cD{\mathcal{D}}
\def\cE{\mathcal{E}}

\def\cH{\mathcal{H}}

\def\sf0{{\mathsf{0}}}

\def\nn{\nonumber}








%% file: digital_twin_csi_AA2.bbl
\begin{thebibliography}{10}
	\providecommand{\url}[1]{#1}
	\csname url@samestyle\endcsname
	\providecommand{\newblock}{\relax}
	\providecommand{\bibinfo}[2]{#2}
	\providecommand{\BIBentrySTDinterwordspacing}{\spaceskip=0pt\relax}
	\providecommand{\BIBentryALTinterwordstretchfactor}{4}
	\providecommand{\BIBentryALTinterwordspacing}{\spaceskip=\fontdimen2\font plus
		\BIBentryALTinterwordstretchfactor\fontdimen3\font minus
		\fontdimen4\font\relax}
	\providecommand{\BIBforeignlanguage}[2]{{%
			\expandafter\ifx\csname l@#1\endcsname\relax
			\typeout{** WARNING: IEEEtran.bst: No hyphenation pattern has been}%
			\typeout{** loaded for the language `#1'. Using the pattern for}%
			\typeout{** the default language instead.}%
			\else
			\language=\csname l@#1\endcsname
			\fi
			#2}}
	\providecommand{\BIBdecl}{\relax}
	\BIBdecl
	
	\bibitem{wen2018deep}
	C.-K. Wen \emph{et~al.}, ``{Deep learning for massive MIMO CSI feedback},''
	\emph{IEEE Wireless Commun. Letters}, vol.~7, no.~5, pp. 748--751, 2018.
	
	\bibitem{wang2018deep}
	T.~Wang \emph{et~al.}, ``{Deep learning-based CSI feedback approach for
		time-varying massive MIMO channels},'' \emph{IEEE Wireless Commun. Letters},
	vol.~8, no.~2, pp. 416--419, 2018.
	
	\bibitem{guo2020convolutional}
	J.~Guo \emph{et~al.}, ``{Convolutional neural network-based multiple-rate
		compressive sensing for massive MIMO CSI feedback: Design, simulation, and
		analysis},'' \emph{IEEE Trans. on Wireless Commun.}, vol.~19, no.~4, pp.
	2827--2840, 2020.
	
	\bibitem{ji2021clnet}
	S.~Ji and M.~Li, ``{CLNet: Complex input lightweight neural network designed
		for massive MIMO CSI feedback},'' \emph{IEEE Wireless Commun. Letters},
	vol.~10, no.~10, pp. 2318--2322, 2021.
	
	\bibitem{cui2022unsupervised}
	Y.~Cui \emph{et~al.}, ``{Unsupervised online learning in deep learning-based
		massive MIMO CSI feedback},'' \emph{IEEE Commun. Letters}, vol.~26, no.~9,
	pp. 2086--2090, 2022.
	
	\bibitem{alkhateeb2023real}
	A.~Alkhateeb, S.~Jiang, and G.~Charan, ``{Real-time digital twins: Vision and
		research directions for 6G and beyond},'' \emph{IEEE Commun. Mag.}, 2023.
	
	\bibitem{zhao2020review}
	S.~Zhao \emph{et~al.}, ``A review of single-source deep unsupervised visual
	domain adaptation,'' \emph{IEEE Trans. on Neural Networks and Learn. Sys.},
	vol.~33, no.~2, pp. 473--493, 2020.
	
	\bibitem{mansour2009domain}
	Y.~Mansour \emph{et~al.}, ``{Domain adaptation: Learning bounds and
		algorithms},'' \emph{arXiv preprint arXiv:0902.3430}, 2009.
	
	\bibitem{robins1995catastrophic}
	A.~Robins, ``Catastrophic forgetting, rehearsal and pseudorehearsal,''
	\emph{Connection Science}, vol.~7, no.~2, pp. 123--146, 1995.
	
	\bibitem{deepmimo}
	A.~Alkhateeb, ``{DeepMIMO}: A generic deep learning dataset for millimeter wave
	and massive {MIMO} applications,'' in \emph{Proc. of Inf. Theory and Appl.
		Workshop}, Feb 2019, pp. 1--8.
	
\end{thebibliography}
